\documentclass[prd,twocolumn,showpacs,floatfix,amsmath,nofootinbib,amssymb,floatfix]{revtex4}
\usepackage{graphicx,color,dcolumn,booktabs,bm,multirow}
\usepackage{longtable,lscape}
\usepackage{txfonts}
 \usepackage{enumitem}
\usepackage{overpic}
\usepackage{amssymb}
\usepackage{indentfirst}
\usepackage{feynmf}   
\usepackage{slashed}  
\usepackage{cases}
\usepackage{color}
\usepackage{multirow}
\usepackage{epstopdf}
\usepackage{graphicx,color,dcolumn,booktabs,bm}
\usepackage[colorlinks,
            citecolor=blue,
            anchorcolor=red,
            menucolor=red,
            linkcolor=red,
            filecolor=red,
            runcolor=red,
            urlcolor=blue,
            frenchlinks=red]{hyperref}
\usepackage{amsmath}
\usepackage{epsfig}
\usepackage{youngtab}
\usepackage{graphicx}
\usepackage{txfonts}

\begin{document}
\title{Investigation of the analog of the $P_{c}$ states-the doubly charmed molecular pentaquarks}

\author{Xuejie Liu$^1$}\email[E-mail: ]{1830592517@qq.com}
\author{Yue Tan$^{2}$}\email[E-mail:]{tanyue@ycit.edu.cn}
\author{Xiaoyun Chen$^{6}$}\email[E-mail:]{xychen@jit.edu.cn}
\author{Dianyong Chen$^{3,4}$\footnote{Corresponding author}}\email[E-mail:]{chendy@seu.edu.cn}
\author{Hongxia Huang$^5$}\email[E-mail:]{hxhuang@njnu.edu.cn}
\author{Jialun Ping$^5$}\email[E-mail: ]{jlping@njnu.edu.cn}
\affiliation{$^1$School of Physics, Henan Normal University, Xinxiang 453007, People's Republic of China}
\affiliation{$^2$School of Mathematics and Physics, Yancheng Institute of Technology, Yancheng, 224051, People's Republic of China}
\affiliation{$^3$School of Physics, Southeast University, Nanjing 210094, People's Republic of China}
\affiliation{$^4$Lanzhou Center for Theoretical Physics, Lanzhou University, Lanzhou 730000, People's Republic of China}
\affiliation{$^5$Department of Physics, Nanjing Normal University, Nanjing 210023, People's Republic of China}
\affiliation{$^6$College of Science, Jinling Institute of Technology, Nanjing 211169, People's Republic of China}

\begin{abstract}
Motivated by the LHCb Collaboration's observation of a doubly charmed tetraquark state $T_{cc}(3875)$, we systematically investigate the existence of doubly charmed pentaquark states using the resonating group method based on the QDCSM framework. The effective potential of the two involved hadrons and the bound state dynamics are included in the present work. Moreover, we have also calculated the scattering phase shifts of open channels by channel coupling to look for possible resonance states. Our estimations indicate that there is a bound state in $I(J^{P})=\frac{3}{2}(\frac{5}{2}^{-})$, with a mass of $4461.7$ MeV. Additionally, five resonance states can be obtained by coupling the open channel, which are $\Xi_{cc}\rho$ and $\Sigma_{c}D^{\ast}$ with $I(J^{P})=\frac{1}{2}(\frac{1}{2}^{-})$, $\Lambda_{c}D^{\ast}$ and $\Sigma_{c}D^{\ast}$ with $I(J^{P})=\frac{1}{2}(\frac{3}{2}^{-})$ and $\Sigma_{c}D^{\ast}$ with $I(J^{P})=\frac{3}{2}(\frac{1}{2}^{-})$respectively. The existence of these predicted doubly charmed pentaquark states needs to be supported by experimental measurements and discoveries. We hope that some experiments can find evidence of these states.

\end{abstract}

\maketitle

\section{\label{sec:introduction}Introduction}

\begin{figure}[t]
\includegraphics[scale=1.5]{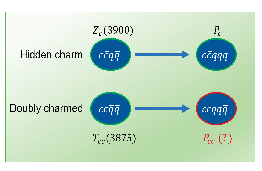}
 \caption{The similarity of the hidden charm and doubly charmed states.}
\label{phase-1.5}
\end{figure}

The discovery of an unexpected charmonium state in 2003 gave birth to the long saga of exotic quarkonia. Since then, numerous charmonium-like states have been experimentally measured, which cannot be directly explained by the baryons with $qqq$ and mesons with $q\bar{q}$. So this new class of hadrons poses a significant challenge to our understanding of the strong interaction. Hence, exploring the structure and properties of these exotic states is a particularly hot topic in hadron physics.

Among the exotic quark states, the $Z_{c}(3900)$ is particularly interesting, as it is the first observed charged charmonium-like state, implying that it contains at least four quarks: $c\bar{c}q\bar{q}$. The $Z_{c}(3900)$ was discovered in 2013 by two independent experiments, BESIII~\cite{BESIII:2013ris} and Belle~\cite{Belle:2013yex}, in the $\pi^{+}J/\psi$ invariant mass spectrum of the process $e^{+}e^{-}\rightarrow \pi^{+}\pi^{-}J/\psi$ at a center of mass energy of 4.26 GeV. Meanwhile, in Ref.~\cite{Xiao:2013iha}, the author analyzed the decay $\psi(4160)\rightarrow J/\psi \pi^{+}\pi^{-}$ and observed the charged $Z_{c}(3900)$ using data of CLEO-c, which reconfirmed the existence of $Z_{c}(3900)$. Besides, further experimental evidence for the $Z_{c}(3900)^{\pm}\rightarrow J/\psi\pi^{\pm}$ came from the semi-inclusive decay of $b-$flavored hadrons in the D0 experiment~\cite{D0:2018wyb}, with the $J/\psi \pi^{+}\pi^{-}$ invariant mass also constrained around the $Y(4260)$ mass region. From these experiments, the favored spin-parity quantum numbers for this peak were $J^{P}=1^{+}$. Extensive theoretical research has been done to understand the nature of the $Z_{c}(3900)$ structure, providing explanations for it as a compact tetraquark~\cite{Braaten:2013boa,Dias:2013xfa,Maiani:2014aja,Qiao:2013raa} or as a $D\bar{D}^{\ast}$ resonance or virtual molecular state~\cite{Albaladejo:2015lob,Wang:2013cya,Wilbring:2013cha,Deng:2014gqa,Guo:2013sya,Dong:2013iqa,Zhang:2013aoa,Aceti:2014uea,Albaladejo:2016jsg,Du:2020vwb,Wang:2020dgr}, or special kinematical near-threshold cusp~\cite{Chen:2013coa,Swanson:2014tra,HALQCD:2016ofq,Pilloni:2016obd}.

The discovery of the $Z_{c}(3900)$ with quark component $c\bar{c}q\bar{q}$ further inspires theoretical physicists to search for similar multiquark states, especially the possible hidden charm pentaquark states. In 2015, the LHCb Collaboration announced the first evidence of two hidden charm pentaquark states $P_{\psi}^{N}(4380)$ and $P_{\psi}^{N}(4450)$ in the $J/\psi p$ invariant mass spectrum measured from the decay process $\Lambda_{b}\rightarrow J/\psi p K^{-}$~\cite{LHCb:2015yax}. Later in the same channel on higher statistics, the LHCb reported the observation of three new pentaquarks: $P_{\psi}^{N}(4312)^{+}$, $P_{\psi}^{N}(4440)^{+}$, and $P_{\psi}^{N}(4457)^{+}$~\cite{LHCb:2019kea}. The obtained hidden charm pentaquark states masses are below and close to the corresponding threshold of the $S-$wave charmed baryon $\Sigma_{c}$ and the $S-$wave anticharmed meson $\bar{D}^{(\ast)}$, which provides the compelling experimental evidence for the existence of $\Sigma_{c}\bar{D}^{\ast}$-type hidden charm molecular pentaquark states in the hadron physics world. Thus, the molecular scheme is one of the more popular explanations~\cite{Wu:2010jy,Wu:2010vk,Wang:2011rga,Yang:2011wz,Yuan:2012wz,Wu:2012md,Garcia-Recio:2013gaa,Xiao:2013yca,Huang:2015uda,Chen:2015loa,Chen:2015moa,Roca:2015dva,He:2015cea,Yang:2015bmv,Meissner:2015mza,Xiao:2015fia,Chen:2016heh,Chen:2016otp,Azizi:2016dhy,Wang:2019got,Guo:2019kdc,Mutuk:2019snd,Zhu:2019iwm,Eides:2019tgv,Weng:2019ynv,Wang:2019nwt,Liu:2019tjn,He:2019ify,Meng:2019ilv,Xiao:2019aya}. However, other interpretations cannot be ruled out, such as the diquark-triquark states~\cite{Lebed:2015tna,Zhu:2015bba}, the diquark-diquark-antiquark states~\cite{Maiani:2015vwa,Anisovich:2015cia,Ghosh:2015xqp,Wang:2015epa}, the genuine multiquark states~\cite{Mironov:2015ica}, the topological soliton~\cite{Scoccola:2015nia}, the kinematical threshold effects in the triangle singularity mechanism~\cite{Guo:2015umn,Liu:2015fea,Mikhasenko:2015vca}, etc., because the structure and properties of the $P_{c}$ states are still unclear.

Furthermore, from the perspective of a multiquark state, the existence of multiquark states with quark content $c\bar{c}q\bar{q}$ implies the possibility of multiquark states with quark content $cc\bar{q}\bar{q}$. Indeed, a new exotic state, $T_{cc}(3875)$, containing two charm quarks, was reported by the LHCb Collaboration in the $D^{0}D^{0}\pi^{+}$ mass distribution~\cite{LHCb:2021vvq,LHCb:2021auc}. The observation of this extremely narrow state has a small binding energy for the $D^{0}D^{\ast+}$ threshold, which stimulated interest in the hadron community. The proximity of the mass of this exotic state to the $D^{0}D^{\ast +}$ and $D^{+}D^{\ast 0}$ thresholds has given support to the molecular picture in these channels~\cite{Dong:2021bvy,Feijoo:2021ppq,Ling:2021bir,Fleming:2021wmk,Ren:2021dsi,Chen:2021cfl,Albaladejo:2021vln,Du:2021zzh,Baru:2021ldu,Santowsky:2021bhy,Deng:2021gnb,Ke:2021rxd}, although a few investigations in the compact tetraquark picture have been accomplished~\cite{Ballot:1983iv,Zouzou:1986qh,Guo:2021yws,Kim:2022mpa,Agaev:2021vur}, even prior to the $T_{cc}(3875)$ discovery~\cite{Ballot:1983iv,Zouzou:1986qh}.

As mentioned above, the relation between $Z_{c}(3900)$ and $P_{c}$ suggests that $T_{cc}(3875)$ should also have a corresponding relation, namely, there should exist multiquark states involving $ccqq\bar{q}$ for $T_{cc}(3875)$. Some earlier investigations on such systems are given in Refs.~\cite{Yan:2018zdt,Dias:2018qhp,Dong:2021bvy,Shimizu:2017xrg,Guo:2017vcf,Liu:2020nil,Chen:2021kad,Chen:2021htr,Shen:2022zvd,Zhou:2018bkn,Park:2018oib,Zhu:2019iwm,Xing:2021yid,Wang:2018lhz}. In Ref.~\cite{Yan:2018zdt}, the authors obtained the masses of the meson-baryon type doubly charmed pentaquark states with $J^{P}=\frac{1}{2}^{-}$ below 4.2 GeV based on the unitarized coupled-channel approach. In Ref.~\cite{Dias:2018qhp}, some deeply bound states of $\Sigma_{c}^{(\ast)}D^{(\ast)}$ with binding energies of about 100 MeV were found through the extension of the chiral unitary approach to describe meson-baryon interactions. Such conclusions are qualitatively consistent with the results of Ref.~\cite{Dong:2021bvy}. Besides, the meson-baryon transitions between the coupled channels  $\Lambda_{c}D^{\ast}$-$\Sigma_{c}^{\ast}D$-$\Sigma_{c}D^{\ast}$-$\Sigma_{C}^{\ast}D^{\ast}$ were performed and a doubly charmed state $\Xi_{cc}^{\ast}(4380)$ exists with almost the same mass as $P_{c}(4380)$ in Ref.~\cite{Shimizu:2017xrg}. Several doubly charmed baryon resonance states can be also found from the $S-$wave scattering of ground states doubly charmed baryons and light pseudoscalar mesons within chiral effective theory~\cite{Guo:2017vcf}. Systematic studies on the $\Sigma_{c}^{(\ast)}D^{(\ast)}$ interaction within the one boson exchange model~\cite{Liu:2020nil,Chen:2021kad}, the chiral effective field theory~\cite{Chen:2021htr} and the unitarized coupled-channel approach~\cite{Shen:2022zvd} were carried out and all the $\Sigma_{c}^{(\ast)}D^{(\ast)}$ systems with isospin $I=\frac{1}{2}$ can be possible doubly charmed molecular pentaquarks with binding energies of about several or dozens MeV, the similar conclusion also can be obtained in Ref.~\cite{Dong:2021bvy}. Furthermore, some compact pentaquark states with doubly heavy quarks were proposed in various models, such as quark model with color magnetic interaction~\cite{Zhou:2018bkn,Park:2018oib}, non-relativistic constituent quark model~\cite{Zhu:2019iwm,Xing:2021yid} and QCD sum rules~\cite{Wang:2018lhz}.

  In the present work, the possible existence of doubly charmed pentaquark states is systematically investigated by using the resonating group method within the framework of the quark delocalization color screening model. The interaction between two involved hadrons with different quantum numbers is studied by evaluating the effective potential. Additionally, the possible bound or resonance states of doubly charmed pentaquarks are estimated by performing the bound state calculation and then taking into account the possible strong decay channels of the doubly charmed pentaquark states.

  The article is arranged as follows. The details of the QDCSM are given in Sec.~\ref{mod}. In Sec.~\ref{dis}, we present the numerical results and discussions, and the last section is reserved for our conclusion.

\par

\section{THE QUARK DELOCALIZATION COLOR SCREENING MODEL  \label{mod}}

The QDCSM is an extension of the native quark cluster model~\cite{DeRujula:1975qlm,Isgur:1979be,Isgur:1978wd,Isgur:1978xj} and was developed with aim of
addressing multiquark systems (More detail of QDCSM can be found in the Refs.~\cite{Wang:1992wi,Chen:2007qn,Chen:2011zzb,Wu:1996fm,Huang:2011kf}).
In the QDCSM, the general form of the Hamiltonian for the pentaquark system is,
\begin{equation}
H = \sum_{i=1}^{5} \left(m_i+\frac{\boldsymbol{p}_i^2}{2m_i}\right)-T_{\mathrm{CM}}+\sum_{j>i=1}^5V(r_{ij}),\\
\end{equation}
where the center-of-mass kinetic energy, $T_{\mathrm{CM}}$, is subtracted without losing generality since we mainly focus on the internal relative motions of the multiquark system. The two body potentials include the color-confining potential, $V_{\mathrm{CON}}$, one-gluon exchange potential, $V_{\mathrm{OGE}}$, and Goldstone-boson exchange potential, $V_{\chi}$, respectively, i.e.,
\begin{equation}
V(r_{ij}) = V_{\mathrm{CON}}(r_{ij})+V_{\mathrm{OGE}}(r_{ij})+V_{\chi}(r_{ij}).
\end{equation}

Noted herein that the potentials include the central, spin-spin, spin-orbit, and tensor contributions, respectively. Since the current calculation is based on S-wave, only the first two kinds of potentials will be considered attending the goal of the present calculation and for clarity in our discussion. In particular, the one-gluon-exchange potential, $V_{\mathrm{OGE}}(r_{ij})$, reads,
\begin{eqnarray}
V_{\mathrm{OGE}}(r_{ij}) &=& \frac{1}{4}\alpha_{ij} \boldsymbol{\lambda}^{c}_i \cdot\boldsymbol{\lambda}^{c}_j \nonumber\\
&&\times\left[\frac{1}{r_{ij}}-\frac{\pi}{2}\delta(\boldsymbol{r}_{ij})\left(\frac{1}{m^2_i}+\frac{1}{m^2_j}
+\frac{4\boldsymbol{\sigma}_i\cdot\boldsymbol{\sigma}_j}{3m_im_j}\right)\right],\ \
\end{eqnarray}
where $m_{i}$ is the quark mass, $\boldsymbol{\sigma}$ and $\boldsymbol{\lambda^{c}}$ are the Pauli matrices and SU(3) color matrices, respectively. The QCD-inspired effective scale-dependent strong coupling constant, $\alpha_{ij}$, offers a consistent description of mesons and baryons from the light to the heavy quark sectors. Their values are associated with the quark flavor and determined by the mass difference of the hadrons. It is worth mentioning that a conventional meson contains only one quark and one antiquark, while a baryon has three quarks, which suggests the existence of three-body interactions in the baryon system. Therefore, when using a simple two-body interaction to reproduce the meson and baryon spectrum in a non-relativistic quark model with OGE potential, the parameter values $\alpha_{qq^\prime}$ and $\alpha_{q\bar{q}^\prime}$, which are determined individually by the baryon and meson spectrum, are not the same.

In the QDCSM, the confining interaction $V_{\mathrm{CON}}(r_{ij})$ can be expressed as
\begin{equation}
 V_{\mathrm{CON}}(r_{ij}) =  -a_{c}\boldsymbol{\lambda^{c}_{i}\cdot\lambda^{c}_{j}}\Big[f(r_{ij})+V_{0_{ij}}\Big] \ ,
\end{equation}
where $a_{c}$ represents the strength of the confinement potential and $V_{0_{ij}}$ refers to the zero-point potential. In the case of quark-quark interactions, the value of $V_{0_{qq}}$ is determined based on the differences between theoretical estimations and experimental measurements of baryon masses. This value is the same for quarks with different flavors. On the other hand, for quark-antiquark interactions, the value of $V_{0_{q\bar{q}}}$ is determined by reproducing the mass differences between theoretical estimations and experimental measurements of the meson masses, which is also flavor-independent. Moreover, in the quark delocalization color screening model, the quarks in the considered pentaquark state $ccnn\bar{n}$ are first divided into two clusters, which is baryon cluster composed of three quark and meson cluster composed of one quark and one antiquark. And then the five-body problem can be simplified as a two-body problem the $f(r_{ij})$ is,
\begin{equation}
 f(r_{ij}) =  \left\{ \begin{array}{ll}r_{ij}^2 & \quad \mbox{if }i,j\mbox{ occur in the same cluster}, \\
\frac{1 - e^{-\mu_{ij} r_{ij}^2} }{\mu_{ij}} & \quad \mbox{if }i,j\mbox{ occur in different cluster},
\end{array} \right.
\label{Eq:fr}
\end{equation}
where the color screening parameter $\mu_{ij}$ is determined by fitting the deuteron properties, nucleon-nucleon and nucleon-hyperon scattering phase shifts~\cite{Chen:2011zzb, Ping:1993me,Wang:1998nk}, with $\mu_{nn}= 0.45\ \mathrm{fm}^{-2}$, $\mu_{ns}= 0.19\ \mathrm{fm}^{-2}$
and $\mu_{ss}= 0.08\ \mathrm{fm}^{-2}$, satisfying the relation $\mu_{ns}^{2}=\mu_{nn}\mu_{ss}$, where $n$ represents $u$ or $d$ quark. From this relation, a fact can be found that the heavier the quark, the smaller the parameter $\mu_{ij}$. When extending to the heavy-quark case, we investigate the mass spectrum of $P_{\psi}^N$ with $\mu_{cc}$ varying from $10^{-4}$ to $10^{-2}$ fm$^{-2}$~\cite{Huang:2015uda}, and our estimation indicated that the dependence of the parameter $\mu_{cc}$ is not very significant\footnote{ The typical size of the multiquark system should be several femtometres, for example, if the size of the multiquark system to be 2 fm, then one can find $\mu_{cc}r^2 \propto (10^{-4} \sim 10^{-2})$,\  $\mu_{cn}r^2 =\sqrt{\mu_{cc}\mu_{nn}} r^2 \propto   (10^{-2}\sim 10^{-1})$ and $\mu_{cs} r^2 =\sqrt{\mu_{cc}\mu_{ss}} r^2 \propto (10^{-2}\sim 10^{-1})$, thus the value of the $\mu_{ij} r^2$ is rather small when at least one charm quark included, and in this case, the exponential function can be approximated to be,
\begin{eqnarray}\label{muij}
  e^{-\mu_{ij}r_{ij}^{2}} &=& 1-\mu_{ij}r_{ij}^{2}+\mathcal{O}\left(\mu_{ij}^2 r_{ij}^4\right).
\end{eqnarray}
Accordingly, the confinement potential between two clusters is approximated to be,
\begin{eqnarray}
  V_{\mathrm{CON}}(r_{ij}) &=&  -a_{c}\boldsymbol{\mathbf{\lambda}}^c_{i}\cdot
\boldsymbol{\mathbf{
\lambda}}^c_{j}~\left(\frac{1-e^{-\mu_{ij}\mathbf{r}_{ij}^2}}{\mu_{ij}}+
V_{0_{ij}}\right) \nonumber \\
  ~ &\approx & -a_{c}\boldsymbol{\mathbf{\lambda}}^c_{i}\cdot
\boldsymbol{\mathbf{ \lambda}}^c_{j}~\left(r_{ij}^2+ V_{0_{ij}}\right),
\end{eqnarray}
which is the same as the expression of two quarks in the same cluster. Thus, when the value of the $\mu_{cc}$ is very small, the screened confinement will return to the quadratic form, which is why the results are insensitive to the value of $\mu_{cc}$.}. In the present work, we take $\mu_{cc}=0.01\ \mathrm{fm}^{-2}$. Then $\mu_{sc}$ and $\mu_{nc}$ are obtained by the relations $\mu_{sc}^{2}=\mu_{ss}\mu_{cc} $ and $\mu_{nc}^{2}=\mu_{nn}\mu_{cc}$, respectively.

The Goldstone-boson exchange interactions between light quarks appear because of the dynamical breaking of chiral symmetry. The following $\pi$, $K$, and $\eta$ exchange terms work between the chiral quark-(anti)quark pair, which read,
\begin{eqnarray}
V_{\chi}(r_{ij}) & =&  v^{\pi}_{ij}(r_{ij})\sum_{a=1}^{3}\lambda_{i}^{a}\lambda_{j}^{a}+v^{K}_{ij}(r_{ij})\sum_{a=4}^{7}\lambda_{i}^{a}\lambda_{j}^{a}+v^{\eta}_{ij}(r_{ij})\nonumber\\
&&\left[\left(\lambda _{i}^{8}\cdot
\lambda _{j}^{8}\right)\cos\theta_P-\left(\lambda _{i}^{0}\cdot
\lambda_{j}^{0}\right) \sin\theta_P\right], \label{sala-Vchi1}
\end{eqnarray}
with
\begin{eqnarray}
  v^{B}_{ij} &=&  {\frac{g_{ch}^{2}}{{4\pi}}}{\frac{m_{B}^{2}}{{\
12m_{i}m_{j}}}}{\frac{\Lambda _{B}^{2}}{{\Lambda _{B}^{2}-m_{B}^{2}}}}
m_{B}     \nonumber    \\
&&\times\left\{(\boldsymbol{\sigma}_{i}\cdot\boldsymbol{\sigma}_{j})
\left[ Y(m_{B}\,r_{ij})-{\frac{\Lambda_{B}^{3}}{m_{B}^{3}}}
Y(\Lambda _{B}\,r_{ij})\right] \right\},
\end{eqnarray}
with $B=(\pi, K,  \eta)$ and $Y(x)=e^{-x}/x$ to be the standard Yukawa function. $\boldsymbol{\lambda^{a}}$ is the SU(3) flavor Gell-Mann matrix. The masses of the $\eta$, $K$ and $\pi$ meson are taken from the experimental value~\cite{ParticleDataGroup:2018ovx}. By matching the pion exchange diagram of the $NN$ elastic scattering process in the quark level and in the hadron level, one can relate the $\pi qq$ coupling with the one of $\pi NN$, which is~\cite{Vijande:2004he,Fernandez:1986zn},
\begin{equation}
\frac{g_{ch}^{2}}{4\pi}=\left(\frac{3}{5}\right)^{2} \frac{g_{\pi NN}^{2}}{4\pi} {\frac{m_{u,d}^{2}}{m_{N}^{2}}},
\end{equation}
which assumes that the flavor SU(3) is an exact symmetry, and only broken by the masses of the strange quark. As for the coupling $g_{\pi NN}$, it was determined by the $NN$ elastic scattering~\cite{Fernandez:1986zn}.

Besides, based on the uncertainties in the parameters of the quark model, with the Minuit program, we can determine a set of optimized parameters with errors by fitting the masses of the ground state mesons and baryons in QDCSM. The model parameters with errors are shown in Table~\ref{biaoge} and the masses of the fitted mesons and baryons are listed in Table~\ref{mass}.

\begin{table}[ht]
\caption{\label{biaoge}The values of the model parameters. The masses of mesons and baryons take their experimental values.  $m_\pi$=0.7 fm$^{-1}$, $m_K$=2.51 fm$^{-1}$, $m_\eta$=2.77 fm$^{-1}$, $\Lambda_{\pi}$=4.2 fm$^{-1}$, $\Lambda_{\eta/K }$=5.2 fm$^{-1}$.    }
\renewcommand\arraystretch{1.5}
\begin{tabular}{p{2.5cm}<\centering p{2.5cm}<\centering p{2.5cm}<\centering }
 \toprule[1pt]
      & Parameter  &Value   \\
      \midrule[1pt]
\multirow{3}{*}{Quark masses}  &$m_u$ (MeV)    & 313 $\pm$ 0.346 \\
              &$m_s$ (MeV)                     & 573 $\pm$ 0.035 \\
              &$m_c$ (MeV)                     & 1788 $\pm$ 0.891\\
      \midrule[1pt]
\multirow{3}{*}{confinement}
              &$a_{c}$ (MeV fm$^{-2}$)         &58.03 $\pm$ 0.589 \\
              &$V_{0_{qq}}$  (fm$^2$)          &-1.2883 $\pm$ 0.001\\
              &$V_{0_{q\bar{q}}}$  (fm$^2$)    &-0.7432 $\pm$ 0.005\\
     \midrule[1pt]
\multirow{5}{*}{OGE}
              &$\alpha_{uu}$                     &0.5652 $\pm$ 0.033\\
              &$\alpha_{uc}$                     &0.2091 $\pm$ 0.025\\
              &$\alpha_{cc}$                     &0.5501 $\pm$ 0.018\\
              &$\alpha_{u\bar{u}}$               &1.4914 $\pm$ 0.003\\
              &$\alpha_{u\bar{c}}$               &0.9629 $\pm$ 0.008 \\
    \midrule[1pt]
Wave function &$b$ (fm)                         &0.518$\pm$ 0.003\\
\bottomrule[1pt]
\end{tabular}
\label{parameters}
\end{table}



\begin{table}[ht]
    \caption{The masses of the ground baryons and mesons in the unit of MeV. Experimental values are taken
        from the Particle Data Group (PDG)~\cite{ParticleDataGroup:2018ovx}.}
    \renewcommand\arraystretch{1.5}
    \begin{tabular}{p{1cm}<\centering p{2.0cm}<\centering  p{1cm}<\centering p{1cm}<\centering p{2cm}<\centering p{1cm}<\centering}
        \toprule[1pt]
        States          &M+$\Delta M$       & PDG     &States        &M+$\Delta M$    & PDG \\
        \midrule[1pt]
         N              &939.1 $\pm$ 19.5   & 939.5   &$\pi$         &139.1 $\pm$ 7.5 & 139.6\\
       $\Delta$         &1232.2 $\pm$ 15.2  & 1232.0  &$\rho$        &770.0 $\pm$ 3.7 & 775.3\\
       $\Xi_{cc}$       &3620.5 $\pm$ 11.2  & 3621.4  &$\omega$      &721.9 $\pm$ 2.8 & 782.7\\
       $\Xi_{cc}^{\ast}$&3632.1 $\pm$ 10.1  & 3696.5  &$\eta$        &283.2 $\pm$ 5.1 & 547.8\\
       $\Lambda_{c}$    &2284.7 $\pm$ 16.6  & 2286.5  &$D^{\ast}$    &1940.2 $\pm$1.4 & 2006.9\\
       $\Sigma_{c}$     &2472.4 $\pm$ 14.2  & 2455.0  & D            &1869.2 $\pm$2.1 & 1869.6\\
       $\Sigma_{c}^\ast$&2483.9 $\pm$ 13.1  & 2520.0                                       \\

        \bottomrule[1pt]
    \end{tabular}
    \label{mass}
\end{table}

In QDCSM, the quark delocalization is realized by specifying the single-particle orbital wave function as a linear combination of left and right Gaussian basis, the single
particle orbital wave functions used in the ordinary quark cluster model reads,
\begin{eqnarray}\label{wave0}
\psi_{\alpha}(\boldsymbol{s}_{i},\epsilon)&=&\left(\Phi_{\alpha}(\boldsymbol{s}_{i})
  +\epsilon\Phi_{\beta}(\boldsymbol{s}_{i})\right)/N(\epsilon), \nonumber \\
\psi_{\beta}(\boldsymbol{s}_{i},\epsilon)&=&\left(\Phi_{\beta}(\boldsymbol{s}_{i})
  +\epsilon\Phi_{\alpha}(\boldsymbol{s}_{i})\right)/N(\epsilon), \nonumber \\
N(\epsilon)&=& \sqrt{1+\epsilon^2+2\epsilon e^{-s^2_{i}/{4b^2}}},\nonumber \\
\Phi_{\alpha}(\boldsymbol{s}_{i})&=&\left(\frac{1}{\pi b^2}\right)^{\frac{3}{4}}
e^{-\frac{1}{2b^2}\left(\boldsymbol{r_\alpha}-\frac{2}{5}s_{i}\right)^2},\nonumber \\
\Phi_{\beta}(-\boldsymbol{s}_{i})&=&\left(\frac{1}{\pi b^2}\right)^{\frac{3}{4}}
e^{-\frac{1}{2b^2}\left(\boldsymbol{r_\beta}+\frac{3}{5}s_{i}\right)^2},
\end{eqnarray}
with $\boldsymbol{s}_{i}$, $i=(1,2,..., n)$, to be the generating coordinates, which are introduced to
expand the relative motion wave function~\cite{Wu:1998wu,Ping:1998si,Pang:2001xx}. The parameter $b$ is indicating the size of the baryon and meson clusters, which is determined by fitting the radius of the baryon and meson by the variational method~\cite{Huang:2018rpb}. In addition, The mixing parameter $\epsilon(s_{i})$ is not an adjusted one but determined variationally by the dynamics of the multi-quark system itself. This assumption allows the multi-quark system to choose its favorable configuration in the interacting process. It has been used to explain the cross-over of the transition between the hadron phase and the quark-gluon plasma phase~\cite{Xu:2007oam, Huang:2011kf}. Due to the effect of the mixing parameter $\epsilon(s_{i})$, there is a certain probability for the quarks between the two clusters to run, which leads to the existence of color octet states for the two clusters. Therefore, this model also includes the hidden color channel effect, which is confirmed by Refs.~\cite{Xia:2021tof,Huang:2020bmb}.

\section{The results and discussions\label{dis}}
In the present work, we investigate the possible lowest-lying and resonance states of the $ ccnn\bar {n}$ (where n=(u or d)) pentaquark systems within QDCSM.
Specifically, we focus on pentaquark states with negative parity, indicating a total angular momentum of zero. Consequently, the total angualr momentum, $J$, aligns with the total spin, $S$, and can have values of $1/2$, $3/2$, and $5/2$. Moreover, all estimation results are based on the molecular state scenario. The possible baryon-meson channels involved which are under consideration in the estimations are listed in Table~\ref{channels}.


\begin{table*}[htb]
\begin{center}
\caption{\label{channels} All possible channels for doubly charmed pentaquark systems with different quantum numbers }
\renewcommand\arraystretch{1.5}
\begin{tabular}{p{1.0cm}<\centering p{1cm}<\centering p{1cm}<\centering p{1cm}<\centering p{1cm}<\centering p{1cm}<\centering p{1cm}<\centering | p{1cm}<\centering p{1cm}<\centering p{1.0cm}<\centering p{1.0cm}<\centering p{1.0cm}<\centering p{1.0cm}<\centering p{1.0cm}<\centering p{1.0cm}<\centering p{1.0cm}<\centering p{1.0cm}<\centering p{1.0cm}<\centering}
\toprule[1pt]
  &    \multicolumn{6}{c|}{$I=\frac{1}{2}$}  & \multicolumn{6}{c}{$I=\frac{3}{2}$}\\
\midrule[1pt]
\multirow{2}{*}{$S=\frac{1}{2}$} &$\Xi_{cc}\eta$ &$\Xi_{cc}\omega$ &$\Xi_{cc}^{\ast}\omega$ &$\Xi_{cc}\pi$ &$\Xi_{cc}\rho$ &$\Xi_{cc}^{\ast}\rho$ &$\Xi_{cc}\pi$ &$\Xi_{cc}\rho$ &$\Xi_{cc}^{\ast}\rho$ &$\Sigma_{c}D$ &$\Sigma_{c}D^{\ast}$ &$\Sigma_{c}^{\ast}D^{\ast}$\\
                                 &$\Lambda_{c}D$ &$\Lambda_{c}D^{\ast}$ &$\Sigma_{c}D$ &$\Sigma_{c}D^{\ast}$ &$\Sigma_{c}^{\ast}D^{\ast}$ &  &\\

\multirow{2}{*}{$S=\frac{3}{2}$} &$\Xi_{cc}\omega$ &$\Xi_{cc}^{\ast}\eta$  &$\Xi_{cc}^{\ast}\omega$  &$\Xi_{cc}\rho$  &$\Xi_{cc}^{\ast}\pi$ &$\Xi_{cc}^{\ast}\rho$  &$\Xi_{cc}^{\ast}\rho$  &$\Xi_{cc}^{\ast}\pi$  &$\Xi_{cc}^{\ast}\rho$  &$\Sigma_{c}D^{\ast}$   &$\Sigma_{c}^{\ast}D$  &$\Sigma_{c}^{\ast}D^{\ast}$\\
&$\Lambda_{c}D^{\ast}$ &$\Sigma_{c}D^{\ast}$  &$\Sigma_{c}D^{\ast}$  &$\Sigma_{c}^{\ast}D^{\ast}$ & & &     \\
$S=\frac{5}{2}$ &$\Xi_{cc}^{\ast}\omega$ &$\Sigma_{c}^{\ast}D^{\ast}$ & & & & &$\Xi_{cc}^{\ast}\pi$  &$\Sigma_{c}^{\ast}D^{\ast}$ \\
\bottomrule[1pt]
\end{tabular}
\end{center}
\end{table*}

\begin{figure}[t]
\includegraphics[scale=0.55]{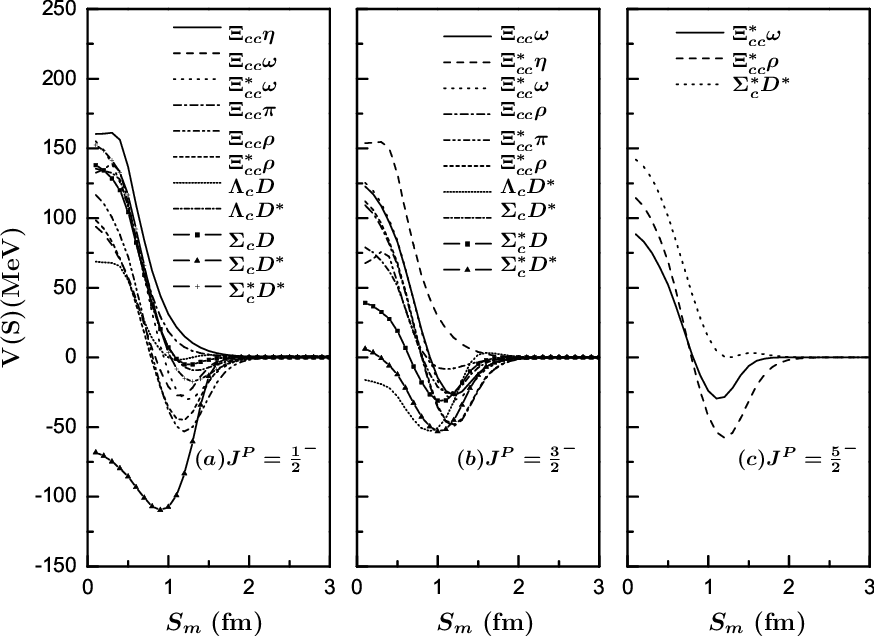}
 \caption{The effective potentials defined in Eq.~(\ref{Eq:PotentialV}) for different channels of the doubly charmed pentaquark systems with $I=1/2$ in QDCSM.}
\label{Veff-0.5}
\end{figure}

\begin{figure}[t]
\includegraphics[scale=0.55]{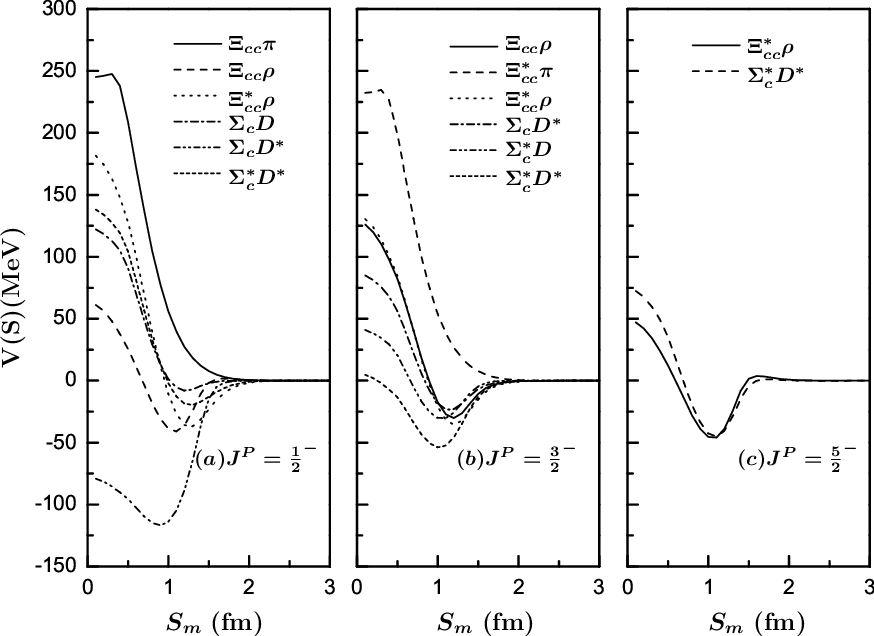}
 \caption{The effective potentials defined in Eq.~(\ref{Eq:PotentialV}) for different channels of the doubly charmed pentaquark systems with $I=3/2$ in QDCSM.}
\label{Veff-1.5}
\end{figure}

\subsection{The effective potentials}
Given that an attractive potential is essential for the formation of a bound state or resonance state, we initially estimate the effective potential between hadron pairs as listed in Table~\ref{channels}. The effective potential is defined as
\begin{eqnarray}
    V(S_m)=E(S_m)-E(\infty), \label{Eq:PotentialV}
\end{eqnarray}
where $E(S_m)$ represents the diagonal matrix element of the system's Hamiltonian in the generator coordinate. Here, $S_{m}$ denotes the separation between the meson and baryon clusters. $E(\infty)$ corresponds to the energy sum of the two clusters when separated by a sufficiently large distance. The effective potentials for doubly charmed pentaquark systems with isospins $I=1/2$ and $I=3/2$ are depicted in Figs.~\ref{Veff-0.5} and \ref{Veff-1.5}, respectively.

For the case of $I(J)^{P}=\frac{1}{2}(\frac{1}{2}^{-})$, there are eleven channels. From Fig.~\ref{Veff-0.5}(a), all the physical channels showattractive interactions expect for $\Xi_{cc}\eta$, $\Xi_{cc}\pi$ and $\Xi_{cc}^{\ast}\omega$. In particular, the attraction between $\Sigma_{c}$ and $D^{\ast}$ is the largest one, followed by that of $\Xi_{cc}\rho$ channel, which is a little larger than of the $\Xi_{cc}\omega$ channel. The attractive interactions of the remaining physical channels are small compared to these channels. These phenomena indicate the $\Sigma_{c}D^{\ast}$, $\Xi_{cc}\rho$ and $\Xi_{cc}\omega$ are more likely to form bound or resonant states due to deeper attractive interactions. For the case of the $I(J)^{P}=\frac{1}{2}(\frac{3}{2}^{-})$, as shown in Fig.~\ref{Veff-0.5}(b), the potential of the $\Xi_{cc}^{\ast}\eta$ channel is repulsive while the potentials of the remaining channel are attractive. From the channels with attractive potentials, $\Lambda_{c}{D}^{\ast}$ and $\Sigma^{\ast}D^{\ast}$ have the largest potentials, which implies that it is possible for $\Lambda_{c}{D}^{\ast}$ and $\Sigma^{\ast}D^{\ast}$ to form bound and resonance states. For the case of $I(J)^{P}=\frac{1}{2}(\frac{5}{2}^{-})$, the potentials are all attractive for channels $\Xi_{cc}^{\ast} \rho$ and $\Xi_{cc}^{\ast}\omega$, while for the $\Sigma_{c}^{\ast}D^{\ast}$ channel, it is strongly repulsive.

In addition, we also estimate the effective potentials of the $I=\frac{3}{2}$ with different angular momentum which are shown in Fig.~\ref{Veff-1.5}.  For the case of $J^{P}=\frac{1}{2}^{-}$, the interaction of $\Xi_{cc}\pi$ channel is repulsive, which means that no bound state or resonance state can be found in this channel, while the potentials of the remaining channels are attractive, so the bound states or resonance states are possible for these channels with attractive potentials. Besides, as shown in Fig.~\ref{Veff-1.5}(a), the attraction between $\Sigma_{c}$ and $D^{\ast}$ is the largest, closely followed by $\Xi_{cc}\rho$ and $\Xi_{cc}^{\ast}\rho$. In addition, the attractions of $\Sigma_{c}^{\ast}D^{\ast}$ and $\Sigma_{c}D$ are the smallest during these five attractive channels. For the case of $J^{P}=\frac{3}{2}^{-}$, similar results to that of $J^{P}=\frac{1}{2}^{-}$ system are obtained. The potentials are all attractive for the $\Sigma_{c}^{\ast}D^{\ast}$, $\Xi_{cc}^{\ast}\rho$, $\Sigma_{c}^{\ast}D$, $\Xi_{cc}\rho$ and $\Sigma_{c}D^{\ast}$ channels expect $\Xi_{cc}^{\ast}\pi$ channels. Furthermore, from Fig.~\ref{Veff-1.5}(b), our estimation also indicates that the attraction potential of $\Sigma_{c}^{\ast}D^{\ast}$ is much stronger than that of the other channels, and the attraction potentials for the remaining channels are similar. For the case of $J^{P}=\frac{5}{2}^{-}$, the effective potentials of $\Xi_{cc}^{\ast}\rho$ and $\Sigma_{c}^{\ast}D^{\ast}$ channels present the property of attractive in Fig.~\ref{Veff-1.5}(c), so one can find that bound states or resonance states are likely to exist due to the attractive nature of the hadron pairs.

\begin{table*}[htb]
\begin{center}
\renewcommand{\arraystretch}{1.5}
\caption{\label{bound-1} The masses of every single channel and those of channel coupling for the pentaquarks with $I=1/2$. The values are provided in units of MeV.}
\begin{tabular}{p{1.2cm}<\centering p{1.2cm}<\centering p{1.6cm}<\centering p{1.6cm}<\centering p{1.8cm}<\centering p{1.6cm}<\centering p{1.6cm}<\centering p{1.8cm}<\centering p{1.6cm}<\centering p{1.6cm}<\centering p{1.2cm}<\centering p{1.2cm}<\centering }
\toprule[1pt]
$I(J^{P})$  &structure   & Channel & $E_{sc}$ &$E_{cc1}$ &$E_{cc2}$   &$E_{sc}^{\prime}$ &$E_{cc1}^{\prime}$ &$E_{cc2}^{\prime}$  &$E_{th}^{Model}$ &$E_{th}^{Exp}$        \\
 \midrule[1pt]
\multirow{11}{*}{$\frac{1}{2}(\frac{1}{2}^{-})$} &\multirow{6}{*}{$QQn-n\bar{n}$} &$\Xi_{cc}\eta$                 &3907.3$\pm$6.2   &3762.5$\pm$3.4  &3759.8$\pm$4.2 &4172.8$\pm$6.2 &3764.5$\pm$3.4 &3764.3$\pm$6.2  &3903.7 &4169.2 \\
                                                 &                                &$\Xi_{cc}\omega$               &4345.3$\pm$13.4  &                &               &4407.1$\pm$13.4 &            & &4342.4 &4404.1\\
                                                 &                                &$\Xi_{cc}^{\ast}\omega$        &4356.6$\pm$12.7  &                &               &4481.8$\pm$12.7 &            & &4353.9 &4479.2\\
                                                 &                                &$\Xi_{cc}\pi$                  &3762.9$\pm$3.7   &                &               &3764.5$\pm$3.7  &            & &3759.5 &3761.0\\
                                                 &                                &$\Xi_{cc}\rho$                 &4389.8$\pm$13.8  &                &               &4396.0$\pm$13.8 &            & &4390.5 &4396.7\\
                                                 &                                &$\Xi_{cc}^{\ast}\rho$          &4403.2$\pm$13.2  &                &               &4473.0$\pm$13.2 &            & &4402.0 &4471.8\\
                                          &\multirow{5}{*}{$Qnn-Q\bar{n}$}        &$\Lambda_{c}D$                 &4157.3$\pm$14.6  &4157.0$\pm$6.2 &               &4159.5$\pm$14.6 &4158.1$\pm$6.2& &4153.9 &4156.1\\
                                                 &                                &$\Lambda_{c}D^{\ast}$          &4228.2$\pm$15.4  &                &               &4296.7$\pm$15.4 &            & &4224.9 &4293.4\\
                                                 &                                &$\Sigma_{c}D$                  &4345.0$\pm$12.3  &                &               &4328.0$\pm$12.3 &            & &4341.6 &4324.6\\
                                                 &                                &$\Sigma_{c}D^{\ast}$           &4370.4$\pm$11.0  &                &               &4419.7$\pm$11.0 &            & &4412.5 &4461.9\\
                                                 &                                &$\Sigma_{c}^{\ast}D^{\ast}$    &4427.3$\pm$11.9  &                &               &4530.2$\pm$11.9 &            & &4424.1 &4526.9\\
\midrule[1pt]
\multirow{10}{*}{$\frac{1}{2}(\frac{3}{2}^{-})$} &\multirow{6}{*}{$QQn-n\bar{n}$} &$\Xi_{cc}\omega$               &4345.3$\pm$13.9  &3773.6$\pm$3.1  &3772.5$\pm$3.2 &4407.0$\pm$13.9 &3838.8$\pm$3.1 &3837.6$\pm$3.2  &4342.4 &4404.1 \\
                                                 &                                &$\Xi_{cc}^{\ast}\eta$          &3918.9$\pm$5.1   &        &       &4247.9$\pm$5.1 &       &        &3915.3 &4244.3\\
                                                 &                                &$\Xi_{cc}^{\ast}\omega$        &4356.7$\pm$12.7  &        &       &4481.9$\pm$12.7 &       &        &4353.9 &4479.2\\
                                                 &                                &$\Xi_{cc}\rho$                 &4390.9$\pm$14.1  &        &       &4397.1$\pm$14.1 &       &        &4390.4 &4396.7\\
                                                 &                                &$\Xi_{cc}^{\ast}\pi$           &3774.1$\pm$2.6   &        &       &3839.2$\pm$2.6 &       &        &3771.0 &3836.1\\
                                                 &                                &$\Xi_{cc}^{\ast}\rho$          &4402.8$\pm$13.1  &        &       &4472.5$\pm$13.1 &       &        &4402.0 &4471.8\\
                                                 &\multirow{4}{*}{$Qnn-Q\bar{n}$} &$\Lambda_{c}D^{\ast}$          &4222.2$\pm$14.9  &4219.9$\pm$14.2&&4290.6$\pm$14.9 &4288.5$\pm$14.2 &        &4224.9 &4293.4\\
                                                 &                                &$\Sigma_{c}D^{\ast}$           &4415.4$\pm$12.9  &        &       &4464.7$\pm$12.9 &       &        &4412.5 &4461.9\\
                                                 &                                &$\Sigma_{c}^{\ast}D$           &4355.5$\pm$11.1  &        &       &4392.0$\pm$11.1 &       &        &4353.1 &4389.6\\
                                                 &                                &$\Sigma_{c}^{\ast}D^{\ast}$    &4421.4$\pm$11.4  &        &       &4524.2$\pm$11.4 &       &        &4424.1 &4526.9\\
\midrule[1pt]
\multirow{3}{*}{$\frac{1}{2}(\frac{5}{2}^{-})$} &\multirow{2}{*}{$QQn-n\bar{n}$} &$\Xi_{cc}^{\ast}\omega$         &4356.9$\pm$12.7  &4402.2$\pm$13.1 &4402.1$\pm$12.5 &4482.2$\pm$12.7 &4471.9$\pm$13.1 &4471.8$\pm$12.5  &4353.9 &4479.2\\
                                                &                                &$\Xi_{cc}^{\ast}\rho$           &4402.2$\pm$12.5  &                &       &4471.9$\pm$12.5 &       &        &4402.0 &4471.8\\
                                                &\multirow{1}{*}{$Qnn-Q\bar{n}$} &$\Sigma_{c}^{\ast}D^{\ast}$     &4425.2$\pm$10.9  &                &       &4527.9$\pm$10.9 &       &        &4424.1 &4526.9\\
\bottomrule[1pt]
\end{tabular}
\end{center}
\end{table*}

\begin{table*}[htb]
\begin{center}
\renewcommand{\arraystretch}{1.5}
\caption{\label{bound-2} The masses of every single channel and those of channel coupling for the pentaquarks with $I=3/2$. The values are provided in units of MeV.}
\begin{tabular}{p{1.2cm}<\centering p{1.2cm}<\centering p{1.6cm}<\centering p{1.6cm}<\centering p{1.8cm}<\centering p{1.6cm}<\centering p{1.8cm}<\centering p{1.8cm}<\centering p{1.6cm}<\centering p{1.6cm}<\centering p{1.2cm}<\centering p{1.2cm}<\centering }
\toprule[1pt]
$I(J^{P})$  &structure   & Channel & $E_{sc}$ &$E_{cc1}$ &$E_{cc2}$   &$E_{sc}^{\prime}$ &$E_{cc1}^{\prime}$ &$E_{cc2}^{\prime}$  &$E_{th}^{Model}$ &$E_{th}^{Exp}$        \\
 \midrule[1pt]
\multirow{6}{*}{$\frac{3}{2}(\frac{1}{2}^{-})$} &\multirow{3}{*}{$QQn-n\bar{n}$} &$\Xi_{cc}\pi$                   &3763.2$\pm$3.7    &3763.2$\pm$3.9  &3763.1$\pm$3.7  &3764.7$\pm$3.7 &3764.7$\pm$3.9 &3764.6$\pm$3.7   &3759.5 &3761.0 \\
                                                &                                &$\Xi_{cc}\rho$                  &4393.1$\pm$14.5   &        &       &4399.3$\pm$14.5 &       &        &4390.5 &4396.7 \\
                                                &                                &$\Xi_{cc}^{\ast}\rho$           &4404.2$\pm$13.4   &        &       &4474.0$\pm$13.4 &       &        &4402.0 &4471.8 \\
                                          &\multirow{3}{*}{$Qnn-Q\bar{n}$}        &$\Sigma_{c}D$                  &4344.9$\pm$12.3   &4344.8$\pm$5.2 &&4328.0$\pm$12.3 &4327.9$\pm$5.2 &        &4341.6 &4324.6 \\
                                                 &                                &$\Sigma_{c}D^{\ast}$           &4363.7$\pm$10.9   &        &       &4413.1$\pm$10.9 &       &        &4412.5 &4461.9 \\
                                                 &                                &$\Sigma_{c}^{\ast}D^{\ast}$    &4427.3$\pm$11.8   &        &       &4530.1$\pm$11.8 &       &        &4424.1 &4526.9 \\
\midrule[1pt]
\multirow{6}{*}{$\frac{3}{2}(\frac{3}{2}^{-})$} &\multirow{3}{*}{$QQn-n\bar{n}$}  &$\Xi_{cc}\rho$                 &4393.3$\pm$14.6  &3774.8$\pm$2.7  &3774.7$\pm$2.7 &4399.6$\pm$14.6 &3839.8$\pm$2.7 &3839.7$\pm$2.7 &4390.5 &4396.7 \\
                                                &                                 &$\Xi_{cc}^{\ast}\pi$           &3774.7$\pm$2.6   &        &       &3839.8$\pm$2.6  &       &       &3771.0 &3836.1 \\
                                                &                                 &$\Xi_{cc}^{\ast}\rho$          &4404.5$\pm$13.4  &        &       &4474.3$\pm$13.4 &       &       &4402.0 &4471.8 \\
                                                &\multirow{3}{*}{$Qnn-Q\bar{n}$}  &$\Sigma_{c}D^{\ast}$           &4415.4$\pm$12.9  &4355.0$\pm$7.1  & &4464.8$\pm$12.9 &4391.5$\pm$7.1 &       &4412.5 &4461.9 \\
                                                &                                 &$\Sigma_{c}^{\ast}D$           &4355.6$\pm$11.0  &        &       &4392.1$\pm$11.0 &       &       &4353.1 &4389.6 \\
                                                &                                 &$\Sigma_{c}^{\ast}D^{\ast}$    &4421.1$\pm$11.4  &        &       &4523.9$\pm$11.4 &       &       &4424.1 &4526.9 \\
\midrule[1pt]
\multirow{2}{*}{$\frac{3}{2}(\frac{5}{2}^{-})$} &\multirow{1}{*}{$QQn-n\bar{n}$} &$\Xi_{cc}^{\ast}\rho$           &4404.4$\pm$13.7  &  &4391.9$\pm$11.2&4474.1$\pm$13.7  & &4461.7$\pm$11.2  &4402.0 &4471.8\\
                                                &\multirow{1}{*}{$Qnn-Q\bar{n}$} &$\Sigma_{c}^{\ast}D^{\ast}$     &4425.5$\pm$11.3  &  &             &4528.3$\pm$11.3  & &                 &4424.1 &4526.9\\
\bottomrule[1pt]
\end{tabular}
\end{center}
\end{table*}

\subsection{Possible bound states}
In this section, to search for possible bound or resonance states, we perform dynamical calculations of the doubly charmed pentaquark systems under the QDCSM employing the resonating group method (RGM)~\cite{Kamimura:1981oxj,Kamimura:1977oxj,Liu:2023oyc}. Here, the doubly charmed pentaquark states are divided into two structures,  $ccn-n\bar{n}$ and $cnn-c\bar{n}$, to be investigated. The estimated results are listed in Table~\ref{bound-1} and~\ref{bound-2}, which correspond to the states with $I=1/2$ and $I=3/2$, respectively. In these tables, $E_{sc}$ is the single channel eigenenergy for different channels; $E_{cc1}$ and $E_{cc2}$ stand for the eigenenergies of the channel-coupling for each kind of structure, and the estimated eigenenergies by simultaneously considering the two structures. $E_{th}^{Model}$ and $E_{th}^{Exp}$ represent the theoretical estimations and experimental measurements of the thresholds of the channels. Considering the uncertainty of the model estimation, we obtain the corrected estimates based on the relative corrective error calculation. As shown in Table~\ref{bound-1} and~\ref{bound-2}, $E_{sc}^{\prime}$, $E_{cc1}^{\prime}$ and $E_{cc2}^{\prime}$ are corrected eigenenergies, and their definitions are the same as those of $E_{sc}$, $E_{cc1}$ and $E_{cc2}$.

As for the $I(J^{P})=\frac{1}{2}(\frac{1}{2}^{-})$ system, the single channel estimations indicate that the obtained central eigenenergies of the $\Sigma_{c}D^{\ast}$ in the $cnn-c\bar{n}$ structure and $\Xi_{cc}\rho$ in the $ccn-n\bar{n}$ structure are lower than the corresponding theoretical threshold of those, and the binding energies are about $-42.1$ MeV and $-0.7$ MeV, respectively. From the Fig.~\ref{Veff-0.5}, the $\Sigma_{c}D^{\ast}$ and $\Xi_{cc}\rho$ have stronger attractive interaction, so this is the reason why two states form bound states.  It is worth the main point that we mainly search for the possible bound states with the mass below the corresponding physical threshold in the present work. In addition, the estimations in Refs.~\cite{Dong:2021bvy,Chen:2021htr,Liu:2020nil,Chen:2021kad,Shen:2022zvd} showed that $\Sigma_{c}D^{\ast}$ may be a good candidate for a doubly charmed molecular state, which accords with that of $\Sigma_{c}D^{\ast}$ single channel estimation.  However, for the remaining channel in the $cnn-c\bar{n}$ structure, along with $\Xi_{cc}\omega$ and $\Xi_{cc}^{\ast}\rho$ in the $ccn-n\bar{n}$ structure, although the effective potentials are attractive, no bound states have been found in those channels because the attractive interactions between the two hadrons are too weak. For the $\Xi_{cc}\eta$, $\Xi_{cc}\pi$ and $\Xi_{cc}^{\ast}\omega$, the obtained central eigenenergies are higher than the corresponding theoretical threshold due to the repulsive nature. After considering the coupling between channels in the same structures, we find that the central eigenenergies are estimated to be 3764.5 MeV and 4158.1 MeV for the $ccn-n\bar{n}$ and $cnn-c\bar{n}$ structures, respectively. Further complete coupled-channels estimations predict a state with a mass to be 3764.3 MeV, which is still above the lowest threshold of $\Xi_{cc}\pi$. In addition, from the estimation of the channel coupling, the obtained central eigenergies are close to that of single channel results, which indicates that the effect of channel coupling is very small in the estimation of the dynamic.

For the $I(J^{P})=\frac{1}{2}(\frac{3}{2}^{-})$ system, there are ten channels in this system as shown in Table~\ref{bound-1}. The single channel estimations show that the $\Lambda_{c}D^{\ast}$ and $\Sigma_{c}^{\ast}D^{\ast}$ in the $cnn-c\bar{n}$ structure are the bound states with the binding energy of about $-2.7$ MeV and $-2.7$ MeV, respectively. It should be noted that the existence of two bound states is due to the deep attractive interaction of $\Lambda_{c}D^{\ast}$ and $\Sigma_{c}^{\ast}D^{\ast}$. Similar conclusions can be drawn in Refs.~\cite{Dong:2021bvy,Chen:2021htr,Liu:2020nil,Chen:2021kad,Shen:2022zvd}, where those results propose the $\Sigma_{c}^{\ast}D^{\ast}$~\cite{Dong:2021bvy,Chen:2021htr,Liu:2020nil,Chen:2021kad,Shen:2022zvd} and $\Lambda_{c}D^{\ast}$~\cite{Dong:2021bvy} with $1/2(3/2^{-})$ to be good candidates for doubly charmed molecular pentaquarks. As for the remaining channel in the two structures except for the $\Xi_{cc}^{\ast}\eta$, there are no bound states because of the weak attraction interaction. For the $\Xi_{cc}^{\ast}\eta$, the potential is repulsive, and thus the obtained central eigenenergy of $\Xi_{cc}^{\ast}\eta$ is also above the corresponding threshold. When considering the channel coupling of the same structure, the obtained central eigenenergy is below the lowest physical channel $\Lambda_{c}D^{\ast}$ threshold with the binding energy of -4.9 MeV in the $cnn-c\bar{n}$ structure while in the $ccn-n\bar{n}$ structure the channel coupling estimation is the opposite of the result for channel coupling under the $cnn-c\bar{n}$ structure. All channels coupling obtain the lowest state whose central energy is $3837.6$ MeV and approach the lowest physical threshold.  For the $I(J^{P})=\frac{1}{2}(\frac{5}{2}^{-})$ system, no pentaquark states below the respective physical thresholds are found in both single channel and multi-channel coupled calculations.

For the $I(J^{P})=\frac{3}{2}(\frac{1}{2}^{-})$ system, a pentaquark molecular state $\Sigma_{c}D^{\ast}$ with a binding energy of about $-48.8$ MeV below its physical threshold is achieved in the dynamical single channel calculation because of the strong attraction between $\Sigma_{c}$ and $D^{\ast}$. In comparison, the central eigenenergies obtained from other single channel calculations surpass their corresponding physical thresholds. Additionally, a comparison of the above results with those of Ref.~\cite{Chen:2021kad} reveals that the conclusion of the bound state calculations for $\Sigma_{c}D^{\ast}$ state with $3/2(1/2^{-})$ is consistent, with both supporting the likelihood of $\Sigma_{c}D^{\ast}$ state being considered a candidate of hadronic molecule. Channel coupling estimation from different structures demonstrates that the central eigenenergies obtained in the $ccn-n\bar{n}$ and $cnn-c\bar{n}$ structures are 3764.7 MeV and 4327.9 MeV, respectively, which are both higher than the physical thresholds of the lowest channels in the relevant structures. The channel coupling calculations for all channels reveal that the central eigenenergy is 3764.6 MeV, which is about 4 above the threshold of the lowest channel $\Xi_{cc}\pi$. For the  $I(J^{P})=\frac{3}{2}(\frac{3}{2}^{-})$ system, the single channel calculation reveals that all the central eigenenergies are above their corresponding physical thresholds, except for $\Sigma_{c}^{\ast}D^{\ast}$, whose central eigenenergy is 4392.1 MeV, which is more than -3.0 MeV below its physical threshold. Neither the $ccn-n\bar{n}$  structure nor the $cnn-c\bar{n}$ structure channel coupling calculation produces any bound states. The central eigenvalue from the channel coupling calculation for all structures is 3839.7 MeV, which is also above the threshold of the lowest channel $\Xi_{cc}^{\ast}\pi$. In the case of $I(J^{P})=\frac{3}{2}(\frac{5}{2}^{-})$ system, it includes two channels, namely $\Xi_{cc}^{\ast}\rho$ and $\Sigma_{c}^{\ast}D^{\ast}$, as shown Table~\ref{bound-2}. Initially, within the single channel estimations, the central eigenenergies for $\Xi_{cc}^{\ast}\rho$ and $\Sigma_{c}^{\ast}D^{\ast}$ are both higher than the physical threshold of their own channels. However, in the channel coupling calculation for all channels, a bound state $\Xi_{cc}^{\ast}\rho$ with a binding energy of -10.1 MeV is found, whose obtained central eigenenergy is 4461.7 MeV.

\begin{figure}[t]
\includegraphics[scale=0.65]{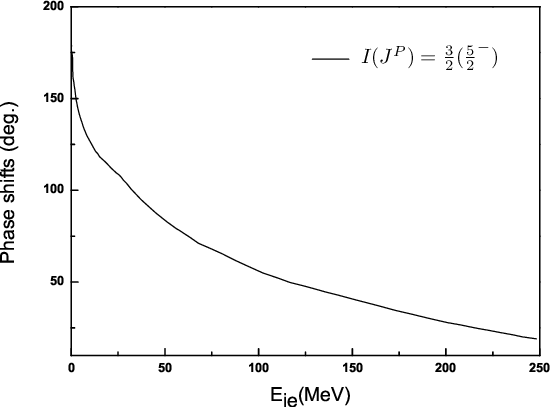}
 \caption{The low-energy scattering phase shifts of the doubly charmed pentaquark systems with $I(J^{P})=\frac{3}{2}(\frac{5}{2}^{-})$ in QDCSM.}
\label{low-phase}
\end{figure}

According to the above estimations, a bound state $\Xi_{cc}^{\ast}\rho$ in $I(J^{P})=\frac{3}{2}(\frac{5}{2}^{-})$ system with a center mass to be 4461.7 MeV is obtained. However, to further check the possibility of this bound state, the low-energy scattering phase shifts of this state are investigated by the variational method, which is shown in Fig.~\ref{low-phase}.
From the Fig.~\ref{low-phase}, the low-energy phase shifts of $\Xi_{cc}^{\ast}\rho$ in the channel coupling estimation can reach up to 180 degrees at $E_{ie}\sim$0 ($E_{ie}$ is the incident energy of the relevant open channel.), and, the valuation of the low-energy phase shifts scattering is intensely reduced at gradually increasing incident energies. This
behavior indicates the presence of a bound state, which is also consistent with the results of the dynamical calculations.

In addition, we have used a new method to calculate the binding energy here. Firstly, based on the low-energy scattering phase shifts calculation, we can obtain the scattering length $a_{0}$ and the effective range $r_{0}$ at the low-energy scattering phase shifts by
  \begin{eqnarray}\label{wave13}
  k \cot{\delta_L} &=& -\frac{1}{a_{0}}+\frac{1}{2}r_{0}k^{2}+O(k^4),
\end{eqnarray}

where $k=\sqrt{2\mu E_{ie}}$, $\mu$, and $E_{ie}$ are the reduced mass of two hadrons and the incident energy, respectively.

According to above results, the wave number $\alpha$ can be available by the relation~\cite{Babenko:2003js},
\begin{eqnarray}\label{wave14}
  r_{0}&=&\frac{2}{\alpha}\left(1-\frac{1}{\alpha a_{0}} \right).
\end{eqnarray}
Finally, the binding energy $E_B^{\prime}$ is calculated according to the relation,
\begin{eqnarray}\label{wave15}
  E_{B}^{\prime}=\frac{\hbar^{2}\alpha^{2}}{2 \mu}.
\end{eqnarray}

\begin{table}[t]
\begin{center}
\renewcommand{\arraystretch}{1.5}
\caption{\label{bound-3} The scattering length $a_{0}$, the effective range $r_{0}$ and the binding energy $E_{B}^{\prime}$ determined by the variation method. }
\begin{tabular}{p{1.5cm}<\centering p{1.5cm}<\centering p{1.8cm}<\centering p{1.8cm}<\centering p{1.8cm}<\centering   }
\toprule[1pt]
$I(J^{P})$  & Channel & $a_{0}$ (fm) &$r_{0}$ (fm) &$E_{B}^{\prime}$ (MeV) \\
\midrule[1pt]
$\frac{3}{2}(\frac{5}{2}^{-})$ & $\Xi_{cc}^{\ast}\rho$       &2.335$\pm$0.092 &0.933$\pm$0.003 & -12.2$\pm$0.004\\
\bottomrule[1pt]
\end{tabular}
\end{center}
\end{table}

For the bound state, the scattering length $a_{0}$, the effective range $r_{0}$, and the binding energy $E_{B}^{\prime}$ are calculated, which are presented in Table~\ref{bound-3}, from which the fact can be seen that the scattering length of $\Xi_{cc}^{\ast}\rho$ is positive, and the binding energy obtained by the variational method, $E_{B}^{\prime}$, is close to the binding energy by the RGM calculations, which further confirm the existence of bound states.

\subsection{Possible resonance states}
\begin{figure*}[t]
\includegraphics[scale=1.0]{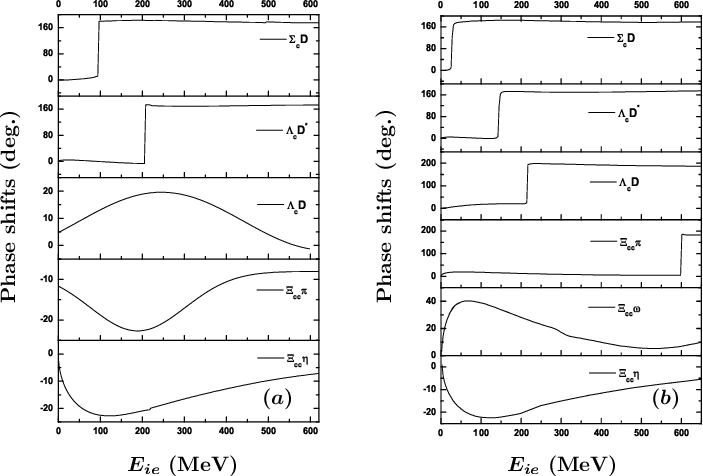}
 \caption{The scattering phase shifts of the open channels with $I(J^{P})=\frac{1}{2}(\frac{1}{2}^{-})$ in QDCSM.    }
\label{phase-0.5-0.5}
\end{figure*}

\begin{figure*}[t]
\includegraphics[scale=1.0]{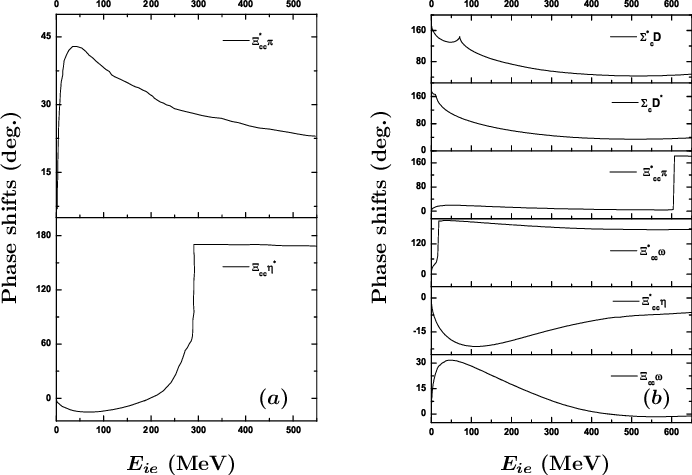}
 \caption{The scattering phase shifts of the open channels with $I(J^{P})=\frac{1}{2}(\frac{3}{2}^{-})$ in QDCSM.}
\label{phase-0.5-1.5}
\end{figure*}

\begin{figure*}[t]
\includegraphics[scale=1.0]{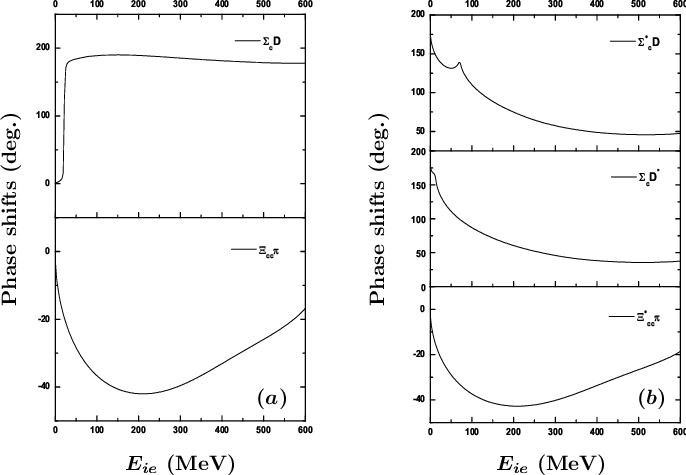}
 \caption{The scattering phase shifts of the open channels with $I=\frac{3}{2}$ in QDCSM.}
\label{phase-1.5}
\end{figure*}


\begin{table*}[htb]
\begin{center}
\renewcommand{\arraystretch}{1.5}
\caption{\label{R} The masses and decay widths (in the unit of MeV) of resonance states with the difference scattering process. $m_R$ stands for the modified resonance mass.  $\Gamma$ is the partial decay width of the resonance state decaying to an open channel. $\Gamma_{Total}$ is the total decay width of the resonance state. }
\begin{tabular}{p{2.8cm}<\centering p{1.cm}<\centering p{1.cm}<\centering p{0.3cm}<\centering p{1.cm}<\centering p{1.cm}<\centering p{0.3cm}<\centering p{1.cm}<\centering p{1.cm} <\centering p{0.3cm} <\centering p{1.cm} <\centering p{1.cm} <\centering p{0.3cm}<\centering p{1.cm} <\centering p{1.cm} <\centering p{1.cm}   }
\toprule[1pt]
                  &\multicolumn{5}{c}{$I(J^{P})=\frac{1}{2}(\frac{1}{2}^{-})$} & &\multicolumn{5}{c}{$I(J^{P})=\frac{1}{2}(\frac{3}{2}^{-})$} & &\multicolumn{2}{c}{$I(J^{P})=\frac{3}{2}(\frac{1}{2}^{-})$}  \\
                  \cline{2-6}\cline{8-12}\cline{14-15}
\multirow{2}{*}{} &\multicolumn{2}{c}{$\Xi_{cc}\rho$} &  &\multicolumn{2}{c}{$\Sigma_{c}D^{\ast}$} & &\multicolumn{2}{c}{$\Lambda_{c}D^{\ast}$} & &\multicolumn{2}{c}{$\Sigma_{c}D^{\ast}$} &  &\multicolumn{2}{c}{$\Sigma_{c}D^{\ast}$}\\
\cline{2-3}\cline{5-6}\cline{8-9}\cline{11-12}\cline{14-15}
\multicolumn{1}{c}{Open channels} &$M_{R}$ &$\Gamma$  & &$M_{R}$ &$\Gamma$ & &$M_{R}$ &$\Gamma$ & &$M_{R}$ &$\Gamma$ & &$M_{R}$ &$\Gamma$\\
 \midrule[1pt]
$\Xi_{cc}\eta$         &$\cdots$   &$\cdots$   &   &$\cdots$     &$\cdots$  &&$\cdots$   &$\cdots$  &&$\cdots$   &$\cdots$ &&$\cdots$   &$\cdots$\\
 $\Xi_{cc}\omega$      &$\cdots$   &$\cdots$   &   &$\cdots$     &$\cdots$  &&$\cdots$   &$\cdots$  &&$\cdots$   &$\cdots$ &&$\cdots$   &$\cdots$\\
 $\Xi_{cc}\pi$         &$\cdots$   &$\cdots$   &   &4461.3       &3.4       &&$\cdots$   &$\cdots$  &&$\cdots$   &$\cdots$ &&$\cdots$   &$\cdots$\\
 $\Lambda_{c}D$        &$\cdots$   &$\cdots$   &   &4420.3       &1.3       &&$\cdots$   &$\cdots$  &&$\cdots$   &$\cdots$ &&$\cdots$   &$\cdots$\\
 $\Lambda_{c}D^{\ast}$ &4391.8     &2.5        &   &4455.3       &3.2       &&$\cdots$   &$\cdots$  &&$\cdots$   &$\cdots$ &&$\cdots$   &$\cdots$\\
 $\Sigma_{c}D$         &4395.5     &1.2        &   &4458.1       &5.1       &&$\cdots$   &$\cdots$  &&$\cdots$   &$\cdots$ &&4431       &5.2     \\
 $\Xi_{cc}^{\ast}\eta$ &$\cdots$   &$\cdots$   &   &$\cdots$   &$\cdots$    &&4286.6     &8.9       &&$\cdots$   &$\cdots$ &&$\cdots$   &$\cdots$\\
 $\Xi_{cc}^{\ast}\pi$   &$\cdots$   &$\cdots$  &   &$\cdots$   &$\cdots$    &&$\cdots$   &$\cdots$  &&4522.2     &1.2      &&$\cdots$   &$\cdots$\\
 $\Xi_{cc}^{\ast}\omega$&$\cdots$   &$\cdots$  &   &$\cdots$   &$\cdots$    &&$\cdots$   &$\cdots$  &&4526.1     &6.4      &&$\cdots$   &$\cdots$\\
 $\Sigma_{c}D^{\ast}$   &$\cdots$   &$\cdots$  &   &$\cdots$   &$\cdots$    &&$\cdots$   &$\cdots$  &&$\cdots$   &$\cdots$ &&$\cdots$   &$\cdots$\\
 $\Sigma_{c}^{\ast}D$   &$\cdots$   &$\cdots$  &   &$\cdots$   &$\cdots$    &&$\cdots$   &$\cdots$  &&$\cdots$   &$\cdots$ &&$\cdots$   &$\cdots$\\
 $\Gamma_{Total}$       &           &3.7       &   &           &13.0        &&           &8.9       &&           &7.6      &&           &5.2\\
\bottomrule[1pt]
\end{tabular}
\end{center}
\end{table*}

From the above bound-state estimation, some bound states are obtained from the single channel calculation due to the strong attraction of two hadrons. However, these states can decay to the corresponding open channels by coupling to some open channels and may be resonance states or scattering states, so here, to check whether the bound states can be transformed into resonance states after coupling to the open channels, the study of the scattering phase shifts of the open channel is needed. It is worth mentioning that only the $S-$wave pentaquark states composed of $ccnn\bar{n}$ are considered in the present work because the width of the high partial waves is almost negligible. Besides, we only consider the two-body decay channels, so the total decay widths of the states given below are the lower limits. We perform the phase shifts of the corresponding open channels, which are shown in Fig.~\ref{phase-0.5-0.5}, ~\ref{phase-0.5-1.5} and ~\ref{phase-1.5}. The mass and width of resonance states are listed in Table.~\ref{R}.

For the $I(J^{P})=\frac{1}{2}(\frac{1}{2}^{-})$ system, two bound states, $\Xi_{cc}\rho$ and $\Sigma_{c}D^{\ast}$, can be found in the single calculation. From the Table~\ref{bound-1}, the bound state $\Xi_{cc}\rho$ can be coupled to five open channel: $\Xi_{cc}\eta$, $\Xi_{cc}\pi$, $\Lambda_{c}D$, $\Lambda_{c}D^{\ast}$ and $\Sigma_{c}D$; the bound state $\Sigma_{c}D^{\ast}$ can be coupled to six open channel: $\Xi_{cc}\eta$, $\Xi_{cc}\omega$, $\Xi_{cc}\pi$, $\Lambda_{c}D$, $\Lambda_{c}D^{\ast}$ and $\Sigma_{c}D$. So the phase shifts of two-channel coupling with a single bound state and the corresponding open channel are shown in Fig.~\ref{phase-0.5-0.5}. From the Fig.~\ref{phase-0.5-0.5}(a), the resonance state $\Xi_{cc}\rho$ can be available in the phase shifts of $\Lambda_{c}D^{\ast}$ and $\Sigma_{c}D$ while there is no resonance state in the phase shifts of $\Xi_{cc}\eta$, $\Xi_{cc}\pi$ and  $\Lambda_{c}D$. For the bound state $\Sigma_{c}D^{\ast}$, one can be seen from the Fig.~\ref{phase-0.5-0.5}(b) that it only has resonance behavior in the scattering phase shifts of $\Lambda_{c}D$, $\Lambda_{c}D^{\ast}$ and $\Sigma_{c}D$. The resonance mass and decay width can be obtained from the shape of the resonance. It is important to note that the x-axes, labeled as $E_{ie}$, in Fig.~\ref{phase-0.5-0.5} represent the incident energy, so the resonance mass $M_{R}$ needs to be acquired by the correction of $M_{R}=E_{ie}(R)-E_{sc}^{\prime}(c)+E_{sc}^{\prime}(o)+E_{th}^{Exp}$, where $E_{ie}(R)$ represents the incident energy at which the resonance phenomenon emerges at $\frac{\pi}{2}$, $E_{sc}^{\prime}(c)$ stands for the bound channel in the single channel calculation, $E_{sc}^{\prime}(o)$ denotes the open channel, and $E_{th}^{Exp}$ means the experimental measurements of the thresholds of the bound channel. Table.~\ref{R} lists the corrected resonance masses and decay widths. The masses of resonance states $\Xi_{cc}\rho$ and $\Sigma_{c}D^{\ast}$ are $(4391.8\sim4395.5)$ MeV and $(4420.3\sim4461.3)$ MeV, respectively, and their decay widths are 3.7 MeV and 13.0 MeV, respectively. From the above estimates, it can be observed that the mass shifts of each resonance state are not larger, which implies that the channel coupling effect between the bound state and the scattering state is weak. This phenomenon is primarily due to the large mass difference between the two channels.

For the $I(J^{P})=\frac{1}{2}(\frac{3}{2}^{-})$ system, the situation is similar to $I(J^{P})=\frac{1}{2}(\frac{1}{2}^{-})$ system. There are also two bound states, $\Lambda_{c}D^{\ast}$ and $\Sigma_{c}D^{\ast}$, in the single-channel calculation. Bound state $\Lambda_{c}D^{\ast}$ can decay to two open channel: $\Xi_{cc}^{\ast}\eta$ and $\Xi_{cc}^{\ast}\pi$, and their scattering phase shifts are shown in Fig.~\ref{phase-0.5-1.5}(a). From Fig.~\ref{phase-0.5-1.5}(a), one can see that there is no resonance state under the scattering phase shifts of $\Xi_{cc}^{\ast}\pi$, but there is a resonance state $\Lambda_{c}D^{\ast}$ in the scattering phase shifts of $\Xi_{cc}^{\ast}\eta$. Its resonance state mass and decay width are $4286.6$ MeV and $8.9$ MeV, respectively. For the bound state $\Sigma_{c}^{\ast}D^{\ast}$, which could be sought in the scattering phase shifts of the open channel: $\Xi_{cc}\omega$, $\Xi_{cc}^{\ast}\eta$, $\Xi_{cc}^{\ast}\omega$, $\Xi_{cc}^{\ast}\pi$, $\Sigma_{c}D^{\ast}$ and $\Sigma_{c}^{\ast}D$, respectively. From Fig.~\ref{phase-0.5-1.5}(b), it can be seen that the resonance state $\Sigma_{c}^{\ast}D^{\ast}$ is only found in the scattering phase shifts of $\Xi_{cc}^{\ast}\omega$ and $\Xi_{cc}^{\ast}\pi$, while the $\Sigma_{c}^{\ast}D^{\ast}$ disappears in the scattering phase shifts of other open channels, which is because the channel coupling effect with $\Sigma_{c}^{\ast}D^{\ast}$ and other open channels pushes the bound state $\Sigma_{c}^{\ast}D^{\ast}$ above its threshold, turning it into a scattering state. From Table~\ref{R}, the resonance mass and decay width are estimated to be $(4522.2\sim4526.1)$ MeV and 7.6 MeV, respectively.

For the $I=\frac{3}{2}$ system, all possible scattering phase shifts of open channels are presented in Fig.~\ref{phase-1.5}. A bound state $\Sigma_{c}D^{\ast}$ is obtained from $J^{P}=\frac{1}{2}^{-}$ though the single calculation, which can decay into open channels: $\Xi_{cc}\pi$ and $\Sigma_{c}D$. By analyzing the scattering phase shifts of two channel coupling with a bound state and a open channel in Fig.~\ref{phase-1.5}(a), the resonance state $\Sigma_{c}D^{\ast}$ can be found in the scattering phase shifts of channel $\Sigma_{c}D$, and the resonance mass and decay width are 4431.0 MeV and 5.2 MeV, respectively. For the $J^{P}=\frac{3}{2}^{-}$ system, a bound state $\Sigma_{c}^{\ast}D^{\ast}$ is shown to exist in the single channel calculation, which can couple with open channels: $\Xi_{cc}^{\ast}\pi$, $\Sigma_{c}D^{\ast}$ and $\Sigma_{c}^{\ast}D$ to perform the scattering phase shifts calculation. A careful study of Fig.~\ref{phase-1.5}(b) reveals the absence of the resonance state in the scattering phase shift of $\Xi_{cc}^{\ast}\pi$, $\Sigma_{c}D^{\ast}$ and $\Sigma_{c}^{\ast}D$. This observation implies that the bound state $\Sigma_{c}^{\ast}D^{\ast}$ has been transformed into a scattering state as a result of the two channel coupling effect.

\section{Summary\label{sum}}

Recently, a tetraquark state $T_{cc}(3875)$ with doubly charmed components, which is below the $D^{\ast+}D^{0}$ mass threshold with $I(J^{P})=1(1^{+})$, was observed by the LHCb Collaboration. These exotic states lead us to wonder: are there doubly charmed pentaquark states in the particle physical world? To explore this question, we systematically estimate the situation for all possible quantum numbers within the framework of QDCSM by using the resonating group method. In the present work, we calculate the effective potentials for each channel to determine whether there is an attractive mechanism between two hadrons, which is a necessary condition for forming a bound state. In addition, we perform dynamic calculations for each single channel and channel coupling. The results show that there is a bound state $\Xi_{cc}\rho$ with quantum number $I(J^{P})=\frac{3}{2}(\frac{5}{2}^{-})$, whose center mass is about 4461.7 MeV.

Furthermore, we also search for possible resonance states by calculating the scattering phase shifts of the open channel. The current estimation results show that five resonance states are present in the doubly charmed pentaquark system, which are $\Xi_{cc}\rho$ and $\Sigma_{c}D^{\ast}$ with $I(J^{P})=\frac{1}{2}(\frac{1}{2}^{-})$ ($M_{R}=4286.6$ MeV, $\Gamma=3.7$ MeV, and $M_{R}=(4420.3\sim4461.3)$ MeV, $\Gamma=13.0$ MeV), $\Lambda_{c}D^{\ast}$ and $\Sigma_{c}D^{\ast}$ with $I(J^{P})=\frac{1}{2}(\frac{3}{2}^{-})$ ($M_{R}=(4391.8-4395.5)$ MeV, $\Gamma=8.9$ MeV, and $M_{R}=(4522.2\sim4526.1)$ MeV, $\Gamma=7.6$ MeV), and $\Sigma_{c}D^{\ast}$ with $I(J^{P})=\frac{3}{2}(\frac{1}{2}^{-})$ ($M_{R}=4431.0$ MeV, $\Gamma=5.2$ MeV).

\acknowledgments{This work is supported partly by the National Natural Science Foundation of China under
Contract Nos. 12175037, 11775118, 11535005, and 11865019; and Jiangsu Provincial Natural Science Foundation Project, No. BK20221166 and National Youth Fund: No. 12205125 also supported this work.}


\end{document}